\documentclass[twocolumn,superscriptaddress,letter]{revtex4}%
\usepackage{amssymb}
\usepackage{color}
\usepackage{graphicx}
\usepackage{dcolumn}
\usepackage{bm}
\usepackage[header,title,page,titletoc]{appendix}
\usepackage{amsmath}
\usepackage{amsfonts}%
\providecommand{\U}[1]{\protect \rule{.1in}{.1in}}
\begin{document}
\title{Non-Hermitian Avalanche Effect -- Non-Perturbative Effect Induced by Local
Non-Hermitian Perturbation on a $Z_{2}$ Topological Order}
\author{Cui-Xian Guo}
\affiliation{Center for Advanced Quantum Studies, Department of Physics, Beijing Normal
University, Beijing 100875, China}
\author{Xiao-Ran Wang}
\affiliation{Center for Advanced Quantum Studies, Department of Physics, Beijing Normal
University, Beijing 100875, China}
\author{Su-Peng Kou}
\thanks{Corresponding author}
\email{spkou@bnu.edu.cn}
\affiliation{Center for Advanced Quantum Studies, Department of Physics, Beijing Normal
University, Beijing 100875, China}

\begin{abstract}
In this paper, based on a non-Hermitian toric-code model, we surprisingly find
that the degeneracy of ground states can be changed by a local non-Hermitian
perturbation (even in thermodynamic limit). We call it non-Hermitian avalanche
effect. As the physics consequences of the non-Hermitian avalanche effect, a
correspondence between bulk quasi-particles and topologically protected
degenerate ground states for $Z_{2}$ topological order is borken down. In addition, the PT symmetry breaking transition of the topologically degenerate ground states subspace can be observed by fidelity susceptibility.

\end{abstract}
\maketitle

\section{Introduction}

Recently, there has been a lot of activities in the research on non-Hermitian
topological
systems\cite{Rudner2009,Esaki2011,Hu2011,Liang2013,Zhu2014,Lee2016,San2016,Leykam2017,Shen2018,Lieu2018,
Xiong2018,Kawabata2018,Gong2018,Yao2018,YaoWang2018,Yin2018,Kunst2018,KawabataUeda2018,Alvarez2018,
Jiang2018,Ghatak2019,Avila2019,Jin2019,Lee2019,Liu2019,38-1,38,chen-class2019,Edvardsson2019,
Herviou2019,Yokomizo2019,zhouBin2019,Kunst2019,Deng2019,SongWang2019,xi2019,Longhi2019,chen-edge2019}%
, including non-Hermitian topological insulators, non-Hermitian topological
superconductors, and non-Hermitian topological semi-metals. After considering
the non-Hermitian extensions of the usual topological band systems, quantum
exotic effects are uncovered, such as the fractional topological invariant and
defective edge states\cite{Lee2016,Yin2018}, non-Hermitian skin
effect\cite{Yao2018,Ghatak2019,Lee2019,SongWang2019,Longhi2019}, and the
breakdown of bulk-boundary
correspondence\cite{Xiong2018,Yao2018,YaoWang2018,Kunst2018,Herviou2019,Yokomizo2019,Kunst2019,Deng2019,Longhi2019}%
. In addition to the research on non-Hermitian topological band systems, the
non-Hermitian extensions of intrinsic topological orders that are many-body
topological systems with long range entanglement are studied\cite{kou-to,ud}.
In Ref.\cite{kou-to}, the non-Hermitian strings and the breakdown of the
correspondence between bulk quasi-particles and topologically protected
degenerate ground states are discovered. In Ref.\cite{ud}, a continuous
quantum phase transition without gap closing was explored that occurs in
non-Hermitian topological orders together with the breakdown of the
Lieb-Robinson bound.

Therefore, one must give it careful reconsideration on the non-Hermitian
extensions of topological stability for intrinsic topological orders. It was
well known that for the topological ordered states, due to the existence of
energy gap, the ground states are robust. The degeneracy of the ground states
depends on the topology of the system and is also robust against any small and
local perturbations. Topological phase transition between topological
ordered states and trivial states may occur when the perturbations become
large enough and are beyond certain thresholds.

In this paper, we will study topological stability for intrinsic topological
orders under non-Hermitian perturbations by taking the non-Hermitian
toric-code model as an example. The effect of non-Hermitian avalanche for a
designed toric-code model is uncovered: for the designed toric-code model with
special external fields, a tiny non-Hermitian perturbation (local imaginary
state selective dissipation) leads to anomalous topological degeneracy and
the breakdown of bulk-degeneracy correspondence (a correspondence between bulk
quasi-particles and topologically protected degenerate ground states).

\section{Topological stability of (Hermitian) $Z_{2}$ topological order}

Firstly, we show the topological stability of (Hermitian) $Z_{2}$ topological order.

For a $Z_{2}$ topological order, there are four types of topological sectors
(ground state and three types of quasi-particles), \textrm{1} (vacuum),
\textrm{e} (e-particle or $Z_{2}$ charge), \textrm{m} (m-particle or $Z_{2}$
vortex), \textrm{f} (fermion). e-particle and m-particle are all bosons with
mutual $\pi$ statistics between them. The fermion can be regarded as a bound
state of an e-particle and an m-particle. All these quasi-particles have
finite energy gaps $\Delta^{I},$ ($I=\mathrm{e},$ \textrm{m}$,$ \textrm{f}).
The four kinds of topological sectors is denoted by $\mathcal{N}=4$ where
$\mathcal{N}$ denotes the number of topological sector of quasi-particles.
When we consider the perturbations on the systems, the energy gaps $\Delta
^{I}$ may change slightly, $\Delta^{I}\rightarrow(\Delta^{I})^{^{\prime}%
}=\Delta^{I}+\delta \Delta^{I}$ ($\delta \Delta^{I}\ll \Delta^{I}$) and cannot be
closed. That indicate all perturbations are irrelevant.

When we consider the system on a torus, the ground states have topological
degeneracy. Each degenerate ground state $\left \vert 0\right \rangle _{I}$
corresponds to the one by adding a virtual quasi-particle. We can use the
basis of sectors of (virtual) quasi-particles to characterize the ground
states, i.e.,
\begin{equation}
\left(  |0\rangle,|e\rangle,|m\rangle,|f\rangle \right)  .
\end{equation}
This is named \emph{bulk-degeneracy correspondence} (BDC). We denote the BDC
by%
\begin{equation}
\mathcal{N}(=4)=\mathcal{D},
\end{equation}
where $\mathcal{D}$ denotes the number of ground state degeneracy. For a
system with infinite size, the four ground states $\left(  |0\rangle
,|e\rangle,|m\rangle,|f\rangle \right)  $ become degenerate with exact zero
energy splitting. When one considers the perturbations on the systems, the
degeneracy of the four ground states doesn't change. In Ref.\cite{zeng}, it
is pointed out that the topological-order classes are stable against any small
stochastic local transformations and there exists a phenomenon of emergence of unitarity.

We use the Kitaev's toric-code model as an example to illustrate the
topological stability of (Hermitian) $Z_{2}$ topological order. The toric-code
model is an exactly solvable spin model, of which the Hamiltonian is
\begin{equation}
\hat{H}_{TC}=-g({\sum_{s}A_{s}}+{\sum_{p}B_{p}}),
\end{equation}
where $A_{s}=\prod_{i\in s}\sigma_{i}^{x}$ and $B_{p}=\prod_{i\in p}\sigma
_{i}^{z}$, the subscripts $s$ and $p$ represent the vertices and
plaquettes of a square lattice, respectively. In this paper, we set $g\equiv1$. For the toric-code model, the
ground states are defined as $A_{s}|\psi_{g}\rangle=|\psi_{g}\rangle,$
$B_{p}|\psi_{g}\rangle=|\psi_{g}\rangle$ for all $A_{s}$ and $B_{p}$.
Furthermore, the elementary excitations are defined as ${A_{s}=-1}$ and
${B_{p}=-1}$.

The quantum states of $Z_{2}$ topological order are characterized by different
configurations of strings, $\hat{W}(C)=\prod_{i\in C}\sigma_{i}^{\alpha_{i}}$
where $\sigma_{i}^{\alpha_{i}}$ is $\alpha_{i}$-type Pauli matrix on site $i$
and $\prod_{i\in C}$ is over all the sites on the string along a loop $C$,
i.e., $\left \vert \Phi \right \rangle =\sum_{C}a_{C}\hat{W}(C)|0\rangle$ where
$|0\rangle$\ denotes the spin polarized states with all spin down ($\left \vert
\downarrow \downarrow,...,\downarrow \right \rangle $), $\hat{W}(C)$ denotes the
possible string operators, and $a_{C}$\ is weight of the string operator. The
different configurations of open strings correspond to different excited
states of different quasi-particles. For e-particle/m-particle, the string
connects the nearest neighboring odd (even) sub-plaquettes
\begin{equation}
\hat{W}_{c/v}(C)=\prod_{i\in C}\sigma_{i}^{s_{c/v}},
\end{equation}
where the product $\prod_{i\in C}$ is over all the sites on the string along a
loop $C$ connecting odd-plaquettes (or even-plaquettes), $s_{c}=z$ and
$s_{v}=x$. The string for f-particles is defined as
\begin{equation}
\hat{W}_{f}(C)=\hat{W}_{v}(C)\hat{W}_{c}(C)=\prod_{i\in C}\sigma_{i}^{s_{f}},
\end{equation}
where $s_{f}=y.$ We point out that the local perturbations on the $Z_{2}$
topological order just locally, and slightly deform the string configurations
but can never change the degeneracy of ground states.

In addition, for the toric-code model, the dissipation effect had been
studied in Ref\cite{nayak}. The results show that small
dissipations cannot change the ground states. As a result, the degenerate
ground states make up a protected code subspace and can be regarded as
topological qubits{ to do possible topological quantum computation}
\cite{kitaev}.

\section{Designed toric-code model and its degenerate ground states}

In this section, we introduce the designed toric-code model, of which the
Hamiltonian is expressed as%
\begin{equation}
\hat{H}_{TC}^{\prime}=\hat{H}_{TC}+\hat{H}^{\prime},
\end{equation}
where
\begin{equation}
\hat{H}^{\prime}=h_{x}\sum_{i\in \mathcal{L}_{1}}\sigma_{i}^{x}+h_{z}\sum
_{i\in \mathcal{L}_{2}}\sigma_{i}^{z}+h_{x}^{\prime}\sum_{i\in \mathcal{L}_{3}%
}\sigma_{i}^{x}.
\end{equation}
Here, $h_{x}$, $h_{z}$ and $h_{x}^{\prime}$ are real parameters, and
$h_{x}^{\prime}$ is a small real parameter.
The dominating external fields are applied only on two crossing lines
($\mathcal{L}_{1}$ and $\mathcal{L}_{2}$). In addition, the auxiliary external
fields are applied on $\mathcal{L}_{3}$. See the illustration in Fig.1.

Under the perturbation $\hat{H}^{\prime}=h_{x}\sum
_{i\in \mathcal{L}_{1}}\sigma_{i}^{x}+h_{z}\sum_{i\in \mathcal{L}_{2}}\sigma
_{i}^{z}+h_{x}^{\prime}\sum_{i\in \mathcal{L}_{3}}\sigma_{i}^{x},$ the
quasi-particles begin to hop. The terms $h_{x}\sum_{i\in \mathcal{L}_{1}}%
\sigma_{i}^{x}$ and $h_{x}^{\prime}\sum_{i\in \mathcal{L}_{3}}\sigma_{i}^{x}$
drives the m-particle without affecting fermion and e-particle along
$\mathcal{L}_{1}$ string and $\mathcal{L}_{3}$ string, respectively. The term
$h_{z}\sum_{i\in \mathcal{L}_{2}}\sigma_{i}^{z}$ drives the e-particle without
affecting fermion and m-particle along $\mathcal{L}_{2}$ string.

\begin{figure}[ptb]
\includegraphics[clip,width=0.5\textwidth]{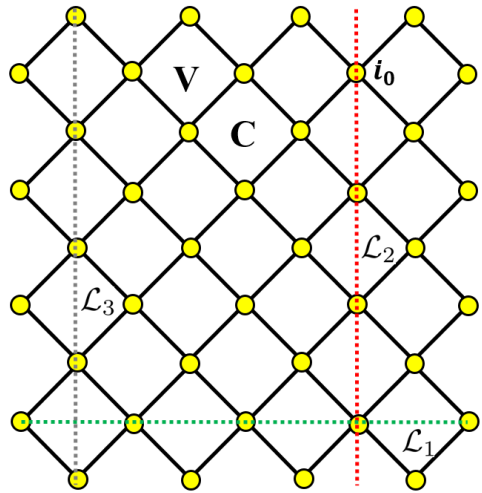}\caption{(Color online)
The schematic diagram of the designed toric-code model. The external fields
are applied only on three paths.}%
\end{figure}

The ground state for $\hat{H}_{TC}$ is a $Z_{2}$ topological
order\cite{kitaev,wen1,kou}. The ground states have topological degeneracy,
i.e., different topologically degenerate ground states are classified by
different topological closed operation strings $\hat{W}_{a}(C
^{\mathrm{close,topo}})$. The operator $\hat{W}_{a}(C^{\mathrm{close,topo}})$
takes on binary values $0,$ $1$ and denotes whether the loops $C
^{\mathrm{close,topo}}$ belong to the even or odd winding number sectors along
the x/y-direction. So, we can use the basis of even-odd parity of the winding
number of electric field lines around the torus $\left \vert m_{ab}%
\right \rangle ,$ $\left(
\begin{array}
[c]{cccc}%
|0,0\rangle & |0,1\rangle & |1,0\rangle & |1,1\rangle
\end{array}
\right)  .$ For a $Z2$ topological order with $4$ degenerate ground states,
there exist the following equations that illustrate the relationship between
the basis of even-odd parity of the winding number of electric field lines
around the torus $\left \vert m_{ab}\right \rangle $ ($a,b=0,1$) and the basis
of topological sectors labeled by different quasi-particles,%
\[
\left(
\begin{array}
[c]{c}%
|0\rangle \\
|e\rangle \\
|m\rangle \\
|f\rangle
\end{array}
\right)  =U\left(
\begin{array}
[c]{c}%
|0,0\rangle \\
|0,1\rangle \\
|1,0\rangle \\
|1,1\rangle
\end{array}
\right)  U^{-1}%
\]
where
\[
U=\frac{1}{\sqrt{2}}\left(
\begin{array}
[c]{cccc}%
1 & 1 & 0 & 0\\
1 & -1 & 0 & 0\\
0 & 0 & 1 & 1\\
0 & 0 & 1 & -1
\end{array}
\right)  .
\]
As a result, the bulk-degeneracy correspondence is valid, i.e.,%
\begin{equation}
\mathcal{N}(=4)=\mathcal{D}.
\end{equation}

We use a four-level system to describe the topologically degenerate ground
states\cite{kou}. After considering $\hat{H}^{\prime},$ three quantum
tunneling processes occur: (1) virtual $Z_{2}$-vortex propagating along
$\mathcal{L}_{1}$ ($\hat{e}_{x}$ direction); (2) virtual $Z_{2}$-charge
propagating along $\mathcal{L}_{2}$ ($\hat{e}_{y}$ direction); (3) virtual
$Z_{2}$-vortex propagating along $\mathcal{L}_{3}$ ($\hat{e}_{y}$ direction)
around the torus. With the help of the high-order purterbative theory, the
four-level quantum system of the four nearly degenerate ground states on a
$2\ast L_{x}\ast L_{y}$ lattice (with $2\ast L_{x}\ast L_{y}$ spins) is
obtained
\begin{equation}
\mathcal{\hat{H}}_{\mathrm{eff}}^{L_{x}\ast L_{y}}=\Delta(\tau_{1}^{x}%
\otimes1)+\varepsilon(\tau_{1}^{z}\otimes \tau_{2}^{x})+\kappa(1\otimes \tau
_{2}^{x}),
\end{equation}
where $\Delta=(\alpha h_{x})^{L_{x}}$, $\varepsilon=(\alpha h_{z})^{L_{y}}$
and $\kappa=(\alpha h_{x}^{\prime})^{L_{y}}$ ($\alpha$ is real parameter). The
eigenvalues of $\mathcal{\hat{H}}_{\mathrm{eff}}^{L_{x}\ast L_{y}}$can be
obtained as $\pm \kappa \pm \sqrt{\Delta^{2}+\varepsilon^{2}}$.

\section{Topological poisoning effect of the non-Hermitian topological order}

We then take the toric-code model as an example to illustrate the string
poisoning effect by considering the non-Hermitian local perturbations. Here,
the non-Hermitian toric-code model is defined by adding non-Hermitian external
fields,
\begin{equation}
\hat{H}_{NTC}^{\prime \prime}=\hat{H}_{TC}+\hat{H}^{\prime \prime},
\end{equation}
where
\begin{equation}
\hat{H}^{\prime \prime}=\sum \limits_{i}\mathbf{h}_{i}\cdot \mathbf{\sigma}%
_{i}=\sum \limits_{i}h_{i}^{x}\sigma_{i}^{x}+\sum \limits_{i}h_{i}^{y}\sigma
_{i}^{y}+\sum \limits_{i}h_{i}^{z}\sigma_{i}^{z}.
\end{equation}
Now, we introduce $\mathbf{h}_{i}\neq \mathbf{h}_{i}^{\ast}$ for $i$th-spin, therefore the Hamiltonian satisfies $\hat{H}_{NTC}^{\prime \prime}\neq \hat{H}_{NTC}^{\prime \prime\ast}$.

To characterize the quantum properties of the non-Hermitian $Z_{2}$
topological order, the (non-Hermitian) dynamic strings were defined
as\cite{kou-to}
\[
D_{a}(C_{N})=\prod_{i\in C}\frac{\hat{t}_{i}^{a}}{\left \vert \hat{t}_{i}%
^{a}\right \vert }=\prod_{i\in C}h_{i}^{a}\sigma_{i}^{a},
\]
where $h_{i}^{a}\sigma_{i}^{a}$ acts at step $i$ for $a$-type excitation and the
indices $a=v,$ $c,$ $f$ correspond to three types of quasi-particles. For the
case of $D_{a}(C_{N})\neq D_{a}^{\dagger}(C_{N}),$ a dynamical string becomes
non-Hermitian. To study its non-Hermitian property, we had introduced the
biorthogonal set for the quantum string states of $Z_{2}$ topological order.

In this paper, we consider the non-Hermitian model with local non-Hermitian
external field on single lattice site $i_{0}$, i.e.,
\begin{equation}
\mathbf{h}_{i_{0}}\neq \mathbf{h}_{i_{0}}^{\ast},\text{ }\mathbf{h}_{i\neq
i_{0}}=\mathbf{h}_{i\neq i_{0}}^{\ast}.
\end{equation}
Now, arbitrary dynamic strings passing through site $i_{0}$ (long or short)
becomes non-Hermitian,
\begin{equation}
D_{a}(C_{N}\rightsquigarrow i_{0})\neq D_{a}^{\dagger}(C_{N}\rightsquigarrow
i_{0}),
\end{equation}
where $C_{N}\rightsquigarrow i_{0}$ means the pathes crossing site $i_{0}.$ We
call it \emph{topological poisoning effect} under local non-Hermitian
perturabtions. Due to the topological poisoning effect, a local
non-Hermitian perturabtion (for example, $\mathbf{h}_{i_{0}}$ at non-Hermitian
external field at site $i_{0}$) may causes highly non-local influence. See the
illustration in Fig.2. The red dashed strings are all non-Hermitian dynamic
strings poisoned by the local non-Hermitian perturbations at site $i_{0}.$

\begin{figure}[ptb]
\includegraphics[clip,width=0.5\textwidth]{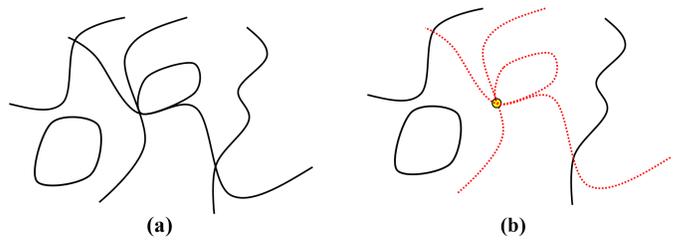}\caption{(Color online)
The schematic diagram of topological poisoning effect: (a) The Hermitian
(dynamic) strings for the Hermitian Z2 topological order; (b) Arbitrary
dynamic strings passing through site $i_{0}$ (long or short) become
non-Hermitian. This is the physics consequence of topological poisoning effect
by adding local non-Hermitian perturbation on site $i_{0},$ i.e., $D_{a}%
(C_{N}\rightsquigarrow i_{0}).$}%
\end{figure}

\section{Non-Hermitian avalanche effect}

\subsection{Local non-Hermitian perturbation}

Now, we consider a particular local non-Hermitian perturbation on the designed
toric-code model,
\begin{equation}
\hat{H}_{NTC}=\hat{H}_{TC}^{\prime}+\hat{H}^{\prime \prime}=\hat{H}_{TC}%
+\hat{H}^{\prime}+\hat{H}^{\prime \prime},
\end{equation}
where
\begin{equation}
\hat{H}^{\prime \prime}=(\lambda_{\mathrm{Re}}+i\lambda_{\mathrm{Im}})
\sigma_{i_{0}}^{z}.
\end{equation}
It is obvious that $\hat{H}_{NTC}$ doesn't have Parity-time symmetry.
However, an important changes is topological poisoning effect under local
non-Hermitian perturabtions, $D_{a}(C_{N}\rightsquigarrow i_{0}).$

When considering above extra non-Hermitian term, the effective Hamiltonian of
the degenerate ground states $\mathcal{\hat{H}}_{\mathrm{eff}}^{L_{x}\ast
L_{y}}$ on designed toric-code model may change:

\begin{enumerate}
\item When the site $i_{0}$ is on vertical dynamic string $\mathcal{L}_{2}$,
$\mathcal{\hat{H}}_{\mathrm{eff}}^{L_{x}\ast L_{y}}$ becomes%
\begin{equation}
\mathcal{\hat{H}}_{\mathrm{eff}}^{L_{x}\ast L_{y}}=\Delta(\tau_{1}^{x}%
\otimes1)+\varepsilon^{\prime}(\tau_{1}^{z}\otimes \tau_{2}^{x})+\kappa
(1\otimes \tau_{2}^{x}),
\end{equation}
where $\Delta=(\alpha h_{x})^{L_{x}}$, $\varepsilon^{\prime}=\alpha^{L_{y}%
}(h_{z})^{L_{y}-1}(h_{z}+\lambda_{\mathrm{Re}}+i\lambda_{\mathrm{Im}})$ and
$\kappa=(\alpha h^{\prime}_{x})^{L_{y}}$ . The eigenvalues of $\mathcal{\hat
{H}}_{\mathrm{eff}}^{L_{x}\ast L_{y}}$ can be obtained as $\pm \kappa \pm
\sqrt{\Delta^{2}+(\varepsilon^{\prime})^{2}}$;

\item When the site $i_{0}$ is on transverse dynamic string $\mathcal{L}_{1}$,
$\mathcal{\hat{H}}_{\mathrm{eff}}^{L_{x}\ast L_{y}}$ becomes%
\begin{equation}
\mathcal{\hat{H}}_{\mathrm{eff}}^{L_{x}\ast L_{y}}=\Delta^{\prime}(\tau
_{1}^{x}\otimes1)+\varepsilon(\tau_{1}^{z}\otimes \tau_{2}^{x})+\kappa
(1\otimes \tau_{2}^{x}),
\end{equation}
where $\Delta^{\prime}=\alpha^{L_{x}}(h_{x})^{L_{x}-1}(h_{x}+\lambda
_{\mathrm{Re}}+i\lambda_{\mathrm{Im}})$, $\varepsilon=(\alpha h_{z})^{L_{y}}$
and $\kappa=(\alpha h^{\prime}_{x})^{L_{y}}$. The eigenvalues of
$\mathcal{\hat{H}}_{\mathrm{eff}}^{L_{x}\ast L_{y}}$ can be obtained as
$\pm \kappa \pm \sqrt{(\Delta^{\prime})^{2}+\varepsilon^{2}}$;

\item When the site $i_{0}$ is on crossing between $\mathcal{L}_{1}$ and
$\mathcal{L}_{2}$, $\mathcal{\hat{H}}_{\mathrm{eff}}^{L_{x}\ast L_{y}}$
becomes%
\begin{equation}
\mathcal{\hat{H}}_{\mathrm{eff}}^{L_{x}\ast L_{y}}=\Delta^{\prime}(\tau
_{1}^{x}\otimes1)+\varepsilon^{\prime}(\tau_{1}^{z}\otimes \tau_{2}^{x}%
)+\kappa(1\otimes \tau_{2}^{x}),
\end{equation}
where $\Delta^{\prime}=\alpha^{L_{x}}(h_{x})^{L_{x}-1}(h_{x}+\lambda
_{\mathrm{Re}}+i\lambda_{\mathrm{Im}})$, $\varepsilon^{\prime}=\alpha^{L_{y}%
}(h_{z})^{L_{y}-1}(h_{z}+\lambda_{\mathrm{Re}}+i\lambda_{\mathrm{Im}})$ and
$\kappa=(\alpha h^{\prime}_{x})^{L_{y}}$. The eigenvalues of $\mathcal{\hat
{H}}_{\mathrm{eff}}^{L_{x}\ast L_{y}}$ can be obtained as $\pm \kappa \pm
\sqrt{(\Delta^{\prime})^{2}+(\varepsilon^{\prime})^{2}}$;

\item When the site $i_{0}$ is not on dynamic strings ($\mathcal{L}_{1}$,
$\mathcal{L}_{2}$ and $\mathcal{L}_{3}$), $\mathcal{\hat{H}}_{\mathrm{eff}%
}^{L_{x}\ast L_{y}}$ doesn't change. The eigenvalues of $\mathcal{\hat{H}%
}_{\mathrm{eff}}^{L_{x}\ast L_{y}}$ can be obtained as $\pm \kappa \pm
\sqrt{\Delta^{2}+\varepsilon^{2}}$.
\end{enumerate}

In this paper, we focus on the case 1 and $\hat{H}^{\prime \prime}%
=-h_{z}+ih_{z}\sigma_{i_{0}}^{x}$. Therefore, The eigenvalues of
$\mathcal{\hat{H}}_{\mathrm{eff}}^{L_{x}\ast L_{y}}$ can be obtained as
$E_{\pm}=\pm \kappa \pm \sqrt{\Delta^{2}-\varepsilon^{2}}$.

\subsection{Spontaneous PT-symmetry breaking}

When the site $i_{0}$ is on vertical dynamic string $\mathcal{L}_{2}$ and $\hat{H}^{\prime \prime}%
=-h_{z}+ih_{z}\sigma_{i_{0}}^{x}$, we have $\varepsilon^{\prime}=i\varepsilon$. Then $\mathcal{\hat{H}%
}_{\mathrm{eff}}^{L_{x}\ast L_{y}}$ becomes%
\begin{equation}
\mathcal{\hat{H}}_{\mathrm{eff}}^{L_{x}\ast L_{y}}= \Delta(\tau_{1}%
^{x}\otimes1)+i\varepsilon(\tau_{1}^{z}\otimes \tau_{2}^{x})+\kappa(1\otimes \tau_{2}^{x}),
\end{equation}
where $\Delta=(\alpha h_{x})^{L_{x}}$, $\varepsilon=(\alpha h_{z})^{L_{y}}$ and $\kappa=(\alpha h^{\prime}_{x})^{L_{y}}$. The effective Hamiltonian
$\mathcal{\hat{H}}_{\mathrm{eff}}^{L_{x}\ast L_{y}}$ has PT symmetry. The
eigenvalues of $\mathcal{\hat{H}}_{\mathrm{eff}}^{L_{x}\ast L_{y}}$ can be
obtained as
\begin{equation}
\begin{split}
 E_{1}&=- \kappa- \sqrt{\Delta
^{2}-\varepsilon^{2}},~~
 E_{2}=- \kappa+ \sqrt{\Delta
^{2}- \varepsilon  ^{2}},\\
 E_{3}&= \kappa- \sqrt{\Delta
^{2}-\varepsilon^{2}},~~
 E_{4}= \kappa+ \sqrt{\Delta
^{2}- \varepsilon  ^{2}}.\\
\end{split}
\end{equation}
When the site $i_{0}$ is on vertical dynamic string, for the case of
$|\Delta|\geq|\varepsilon|$, the system belongs to a phase with $\mathcal{PT}$
symmetry, of which $E$ are real and the eigenvectors are eigenstates of the
symmetry operator, i.e., $\mathcal{PT}|\varphi_{i}\rangle=|\varphi_{i}\rangle
$. For the case of $|\Delta|<|\varepsilon|$, $E$ are complex, and
$\mathcal{PT}|\varphi_{i}\rangle \neq|\varphi_{i}\rangle$. A $\mathcal{PT}%
$-symmetry-breaking transition occurs at the exceptional points $|\Delta
|=|\varepsilon|$, which leads to the following relation $h_{x}=h_{z}$ when
$L_{x}=L_{y}$. It is clear that $|\psi_{1}\rangle$ and $|\psi_{2}\rangle$
compose a pair of $\mathcal{PT}$-symmetry, and $|\psi_{3}\rangle$ and
$|\psi_{4}\rangle$ compose another pair of $\mathcal{PT}$-symmetry.

The eigenstates $|\varphi_{i}\rangle$ of $\mathcal{\hat{H}}_{\mathrm{eff}%
}^{L_{x}\ast L_{y}}$ can be written as
\begin{equation}\label{eqfi}
|\psi_{1}\rangle=\frac{1}{\mathrm{N}_{1}}\left(
\begin{array}
[c]{c}%
\frac{i\varepsilon+\sqrt{\Delta^{2}-\varepsilon^{2}}}{\Delta}\\
\frac{-\Delta^{2}+(\varepsilon-i\sqrt{\Delta^{2}-\varepsilon^{2}}%
)(\varepsilon-i\kappa)}{\Delta(\sqrt{\Delta^{2}-\varepsilon^{2}}+\kappa)}\\
-1\\
1
\end{array}
\right)  ,\nonumber
\end{equation}%
\begin{equation}
|\psi_{2}\rangle=\frac{1}{\mathrm{N}_{2}}\left(
\begin{array}
[c]{c}%
\frac{i\varepsilon-\sqrt{\Delta^{2}-\varepsilon^{2}}}{\Delta}\\
\frac{\Delta^{2}+(-\varepsilon-i\sqrt{\Delta^{2}-\varepsilon^{2}}%
)(\varepsilon-i\kappa)}{\Delta(\sqrt{\Delta^{2}-\varepsilon^{2}}-\kappa)}\\
-1\\
1
\end{array}
\right)  ,\nonumber
\end{equation}%
\begin{equation}
|\psi_{3}\rangle=\frac{1}{\mathrm{N}_{3}}\left(
\begin{array}
[c]{c}%
\frac{i\varepsilon-\sqrt{\Delta^{2}-\varepsilon^{2}}}{\Delta}\\
\frac{-\Delta^{2}+(\varepsilon+i\sqrt{\Delta^{2}-\varepsilon^{2}}%
)(\varepsilon-i\kappa)}{\Delta(\sqrt{\Delta^{2}-\varepsilon^{2}}-\kappa)}\\
1\\
1
\end{array}
\right)  ,\nonumber
\end{equation}
and
\begin{equation}
|\psi_{4}\rangle=\frac{1}{\mathrm{N}_{4}}\left(
\begin{array}
[c]{c}%
\frac{i\varepsilon+\sqrt{\Delta^{2}-\varepsilon^{2}}}{\Delta}\\
\frac{\Delta^{2}+(-\varepsilon+i\sqrt{\Delta^{2}-\varepsilon^{2}}%
)(\varepsilon-i\kappa)}{\Delta(\sqrt{\Delta^{2}-\varepsilon^{2}}+\kappa)}\\
1\\
1
\end{array}
\right)  ,
\end{equation}
where $\mathrm{N}_{i}$ ($i=1,2,3,4$) are normalization constant. In the region
of $\mathcal{PT}$-unbroken phase ($|\Delta|\geq|\varepsilon|$), the
normalization constant are obtained as $\mathcal{N}_{1}=\mathcal{N}%
_{2}=\mathcal{N}_{3}=\mathcal{N}_{4}=2$. And in the region of $\mathcal{PT}%
$-broken phase ($|\Delta|<|\varepsilon|$), the normalization constant are
obtained as $\mathcal{N}_{1}=\mathcal{N}_{4}=\frac{2\sqrt{\varepsilon
^{2}+\varepsilon \sqrt{\varepsilon^{2}-\Delta^{2}}}}{\Delta}$, $\mathcal{N}%
_{2}=\mathcal{N}_{3}=\frac{2\sqrt{\varepsilon^{2}-\varepsilon \sqrt
{\varepsilon^{2}-\Delta^{2}}}}{\Delta}$.

From the result, one can see there exist exceptional points (EPs) at
$|\Delta|=|\varepsilon|$. In the limit of $\Delta \rightarrow0,$ $\varepsilon
\rightarrow0$ according to the condition of quantum phase transition
($|\Delta|=|\varepsilon|$), an arbitrary small local perturbation (a local
complex external field) causes the quantum phase transition for the ground
states. We call it \emph{non-Hermitian avalanche effect}. In the followings,
we show the physics consequences of the non-Hermitian avalanche effect -
breakdowns of bulk-degeneracy correspondence for Z2 topological order.

\subsection{Breakdowns of bulk-degeneracy correspondence for Z2 topological
order}

To characterize the non-Hermitian avalanche effect, we define the
(non-Hermitian) degeneracy $\mathcal{D}$ under local non-Hermitian
perturbation.

Firstly, we define the overlap of any two of these four nearly degenerate
eigenstates as follows,
\begin{equation}\label{eqoo}
\begin{split}
O_{12}  &  =|\langle \psi_{2}|\psi_{1}\rangle|,~O_{13}=|\langle \psi_{3}%
|\psi_{1}\rangle|,~O_{14}=|\langle \psi_{4}|\psi_{1}\rangle|,\\
O_{23}  &  =|\langle \psi_{3}|\psi_{2}\rangle|,~O_{24}=|\langle \psi_{4}%
|\psi_{2}\rangle|,~O_{34}=|\langle \psi_{4}|\psi_{3}\rangle|.
\end{split}
\end{equation}
By inserting Eq.(\ref{eqfi}) into Eq.(\ref{eqoo}) , we obtain the overlap as
\begin{equation}%
\begin{split}
O_{12}  &  =|\frac{\varepsilon}{\Delta}|,~O_{34}=|\frac{\varepsilon}{\Delta
}|,\\
O_{13}  &  =0,~O_{23}=0,~O_{24}=0,~O_{14}=0,
\end{split}
\end{equation}
in the region of $\mathcal{PT}$-unbroken phase ($|\Delta|\geq|\varepsilon|$),
and
\begin{equation}%
\begin{split}
O_{12}  &  =|\frac{\Delta}{\varepsilon}|,~O_{34}=|\frac{\Delta}{\varepsilon
}|,\\
O_{13}  &  =0,~O_{23}=0,~O_{24}=0,~O_{14}=0,
\end{split}
\end{equation}
in the region of $\mathcal{PT}$-broken phase ($|\Delta|<|\varepsilon|$).

Then, we define the degeneracy of ground states in this case as $\mathcal{D}=4-O_{12}-O_{34}$.
According to the results in Fig.3, the degeneracy becomes change under the
non-Hermitian strength. As a result, the
bulk-degeneracy correspondence is broken, i.e.,
\begin{equation}
\mathcal{N}(=4)\neq \mathcal{D}(=4-O_{12}-O_{34}).
\end{equation}

\begin{figure}[ptb]
\includegraphics[clip,width=0.35\textwidth]{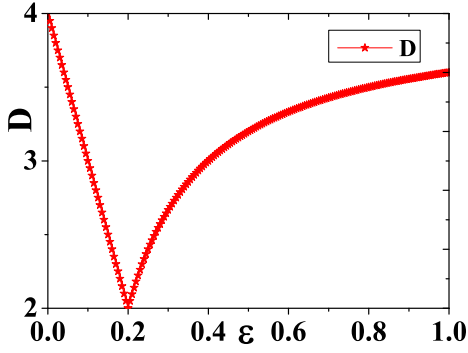}\caption{(Color online)
The non-Hermitian degeneracy that is away from 4. At exceptional point, it is
2.}%
\end{figure}

\subsection{Fidelity susceptibility of ground state}

To confirm the existence of the quantum phase transition from non-Hermitian
avalanche effect, we calculate the fidelity susceptibility of ground state.\

Fidelity susceptibility of ground state can be used to characterize the occurrence of the
quantum phase transitions. In this section, we study fidelity susceptibility of a given
ground state $|\psi_{i}\rangle (i=1,2,3,4)$ in non-Hermitian toric-code model. The
fidelity of ground state in terms of $\varepsilon$ can be defined as
\begin{equation}
F(\varepsilon,\delta)=|\langle \psi_{i}(\varepsilon)|\psi_{i}(\varepsilon
+\delta)\rangle|.
\end{equation}
The fidelity susceptibility of ground state in terms of $\varepsilon$ can be
defined as
\begin{equation}
\chi(\varepsilon,\delta)=\mathrm{lim}_{(\delta \rightarrow0)}\frac{-2\mathrm{ln}%
F}{\delta^{2}}.
\end{equation}
The behavior of $|\psi_{i}\rangle$ of the effective model $\mathcal{\hat{H}%
}_{\mathrm{eff}}^{L_{x}\ast L_{y}}=\Delta(\tau_{1}^{x}\otimes1)+\varepsilon
(\tau_{1}^{z}\otimes \tau_{2}^{x})$ $(\kappa \rightarrow0)$ is same as that of
$H=\Delta \tau_{1}^{x}+\varepsilon \tau_{1}^{z}.$ As a result, the fidelity and
fidelity susceptibility of each ground state are obtained as
\begin{equation}
F(\varepsilon,\delta)=\left \{
\begin{array}
[c]{ll}%
1-\frac{\delta^{2}}{8(\Delta^{2}-\varepsilon^{2})}, &
\hbox{$|\Delta| \geq |\varepsilon|$}\\
1-\frac{\Delta^{2}\delta^{2}}{8\varepsilon^{2}(\varepsilon^{2}-\Delta^{2})}, &
\hbox{$|\Delta| < |\varepsilon|$}
\end{array}
\right.  ,
\end{equation}
and
\begin{equation}
\chi(\varepsilon,\delta)=\left \{
\begin{array}
[c]{ll}%
\frac{1}{4(\Delta^{2}-\varepsilon^{2})}, &
\hbox{$|\Delta| \geq |\varepsilon|$}\\
\frac{\Delta^{2}}{4\varepsilon^{2}(\varepsilon^{2}-\Delta^{2})}, &
\hbox{$|\Delta| < |\varepsilon|$}
\end{array}
\right.  .
\end{equation}

In Fig.4, we plot the fidelity susceptibility of ground states in terms of $\varepsilon$.

\begin{figure}[ptb]
\includegraphics[clip,width=0.35\textwidth]{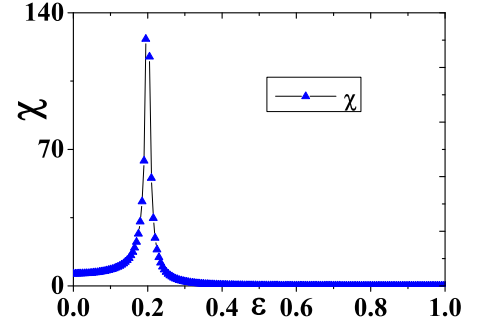}\caption{(Color online)
The fidelity susceptibility of ground states in terms of $\varepsilon$.}
\end{figure}

\subsection{Numerical calculations of the non-Hermitian toric-code model}

To support our theoretical predictions, we do numerical calculations based on
the non-Hermitian toric-code model $\hat{H}_{NTC}$ on $2\ast2\ast2$lattice
and on $2\ast3\ast3$ lattice.

In Fig.5, we plot the numerical results from the exact diagonalization
technique of the non-Hermitian toric code model $\hat{H}_{NTC}$ on $2\ast
2\ast2$ lattice with periodic boundary conditions. Fig.5 shows the global
phase diagram of $\mathcal{PT}$-symmetry-breaking transition for topologically
degenerate ground states. The phase boundary are all exceptional points
characterized by the relation $(\alpha h_{x})^{2}=(\alpha h_{z})^{2}$.

\begin{figure}[ptb]
\includegraphics[clip,width=0.35\textwidth]{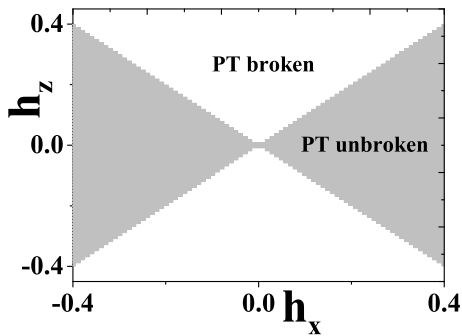}\caption{(Color online)
Phase diagram for spontaneous PT-symmetry breaking for the topologically
degenerate ground states on $2\ast2\ast2$ lattice: in white regions,
PT-symmetry is broken; in the dark regions, PT-symmetry is not broken. The
phase boundaries are exceptional points.}%
\end{figure}

In Fig.6, we plot the numerical results from the exact diagonalization
technique of the non-Hermitian toric-code model $\hat{H}_{NTC}$ on $2\ast
3\ast3$ lattice with periodic boundary conditions. Fig.6 shows the real part
and imaginary part of energy for the four nearly degenerate ground states for
the non-Hermitian toric-code model with $h_{x}=0.1$ and $h'_{x}=0.1$ on $2\ast3\ast3$lattice,
respectively. The numerical results indicate that exceptional points occur
when $h_{x}=h_{z}$, which is consistent with theoretical prediction.

\begin{figure}[ptb]
\includegraphics[clip,width=0.48\textwidth]{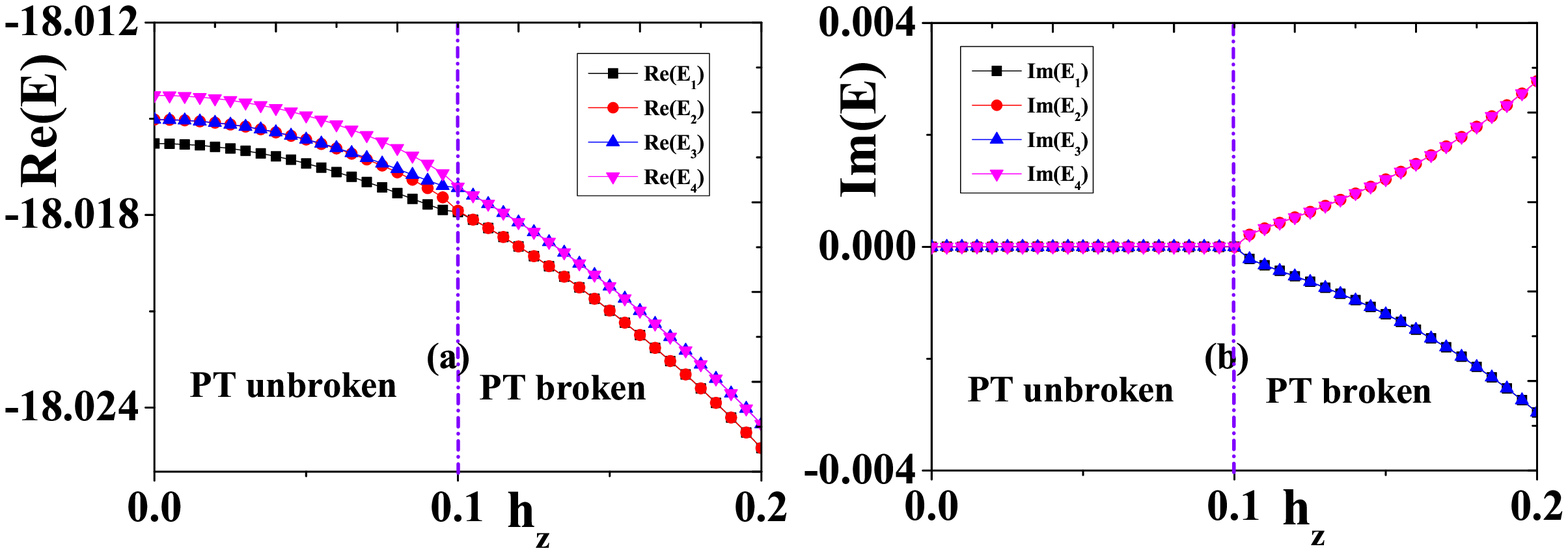}\caption{(Color online)
(a) The real part of energy for the four degenerate ground states for the case
of $h_{x}=0.1$ and $h'_{x}=0.1$ via $h_{z}$ based on the non-Hermitian toric-code model on
$2\ast3\ast3$ lattice; (b) The imaginary of energy for the four degenerate
ground states for the case of $h_{x}=0.1$ and $h'_{x}=0.1$ via $h_{z}$ based on the
non-Hermitian toric-code model on $2\ast3\ast3$ lattice.}%
\end{figure}

In addition, we calculate the overlap of any two of these four nearly
degenerate eigenstates $O_{ij}$defined as above. We theoretically
predict that the overlaps are
\begin{equation}%
\begin{split}
O_{12}  &  =O_{34}=|\frac{\varepsilon}{\Delta}|\sim \Bigg \{
\begin{array}
[c]{c}%
\frac{h_{z}^{2}}{h_{x}^{2}}~~(2\ast2\ast2 ~\mathrm{lattice})\\
\\
\frac{h_{z}^{3}}{h_{x}^{3}}~~(2\ast3\ast3~\mathrm{lattice})
\end{array}
\\
O_{13}  &  =O_{23}=O_{24}=O_{14}=0,
\end{split}
\end{equation}
in $\mathcal{PT}$-unbroken phase. In addition,
\begin{equation}%
\begin{split}
O_{12}  &  =O_{34}=|\frac{\Delta}{\varepsilon}|\sim \Bigg \{
\begin{array}
[c]{c}%
\frac{h_{x}^{2}}{h_{z}^{2}}~~(2\ast2\ast2 ~\mathrm{lattice})\\
\\
\frac{h_{x}^{3}}{h_{z}^{3}}~~(2\ast3\ast3~\mathrm{lattice})
\end{array}
\\
O_{13}  &  =O_{23}=O_{24}=O_{14}=0,
\end{split}
\end{equation}
in $\mathcal{PT}$-broken phase.

In Fig.7, we present the numerical results from the exact diagonalization
technique of the non-Hermitian toric-code model on $2\ast2\ast2$ and
$2\ast2\ast3$ lattices with periodic boundary conditions, respectively. We
plot the the non-Hermitian degeneracy as a function of $h_{z}$ for the case of
$h_{x}=0.1$ and $h'_{x}=0.1$, which is consistent with the theoretical prediction. The results
indicate the degeneracy of ground states may be different from $4$. Now, he
bulk-degeneracy correspondence is broken, i.e.,%
\begin{equation}
\mathcal{N}(=4)\neq \mathcal{D}.
\end{equation}

\begin{figure}[ptb]
\includegraphics[clip,width=0.4\textwidth]{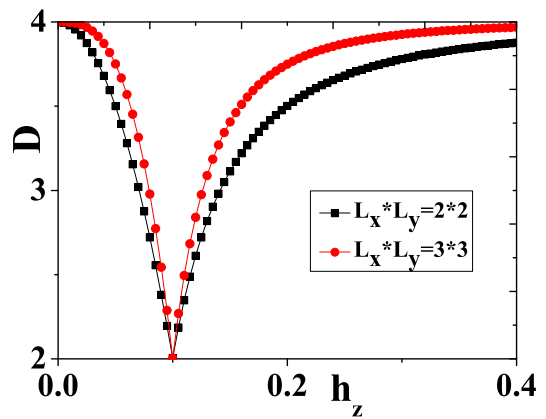}\caption{(Color online)
The non-Hermitian degeneracy for the ground states for the case of $h_{x}=0.1$ and $h'_{x}=0.1$
via $h_{z}$ based on the non-Hermitian toric-code model on $2\ast2\ast2$
lattice and those on $2\ast3\ast3$ lattice.}%
\end{figure}

\begin{figure}[ptb]
\includegraphics[clip,width=0.4\textwidth]{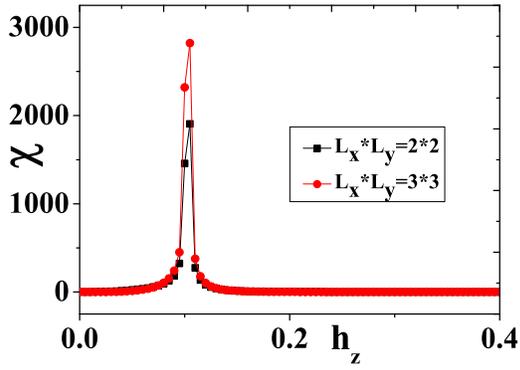}\caption{(Color online)
The fidelity susceptibility for the ground states for the case of $h_{x}=0.1$ and $h'_{x}=0.1$
via $h_{z}$ based on the non-Hermitian toric-code model on $2\ast2\ast2$
lattice and those on $2\ast3\ast3$ lattice.}%
\end{figure}

In Fig.8, we show the fidelity susceptibility of the ground state from the
exact diagonalization technique of the non-Hermitian toric-code model on
$2\ast2\ast2$ and $2\ast2\ast3$ lattices with periodic boundary conditions,
respectively. The results show that the quantum PT phase transition occurs at EPs.

\section{Conclusion}

In this paper, we study the non-Hermitian avalanche effect induced by a local non-Hermitian perturbation. we investigate the effective models for topological degenerate
ground states of the designed non-Hermitian toric-code model by high-order degenerate
perturbation theory. In particular, there exists spontaneous $\mathcal{PT}$-
symmetry breaking for the topologically degenerate ground states subspace. At
\textquotedblleft exceptional points\textquotedblright, the topological
degenerate ground states merge and the topological degeneracy turns into
non-Hermitian degeneracy. Therefore, based on a non-Hermitian toric-code
model, we surprisingly find that the degeneracy of ground states can be
reduced by a local Non-Hermitian perturbation. In addition, the $\mathcal{PT}$-
symmetry breaking transition can be observed by fidelity susceptibility. In the end, the
influence of non-Hermitian on topological order are discussed.

\end{document}